# The Effect of Gravitational Settling on Concentration Profiles and Dispersion Within and Above Fractured Media

Tomer Duman[1,2], Ran Holtzman[3] and Uri Shavit[1,*]

[1] Civil and Environmental Engineering, Technion, Haifa, Israel

[2] Department of biological sciences, Rutgers University, NJ, USA

[3] Department of Soil and Water Sciences, The Hebrew University of Jerusalem, Rehovot, Israel

*Corresponding Author (email: aguri@technion.ac.il, Tel: +972-4-829-3568)

**ABSTRACT**

The transport of heavy particles in a medium that consists of fluid and solid phases such as stream gravel beds, cracked soils and wetlands is affected by processes such as attachment-detachment, gravity and drag, and by mixing processes that are induced by Taylor dispersion and mechanical dispersion. This paper addresses an additional dispersion mechanism which is induced by gravitational settling and is a result of the coupling between the modified particle concentration (the change of particle number density due to settling) and the lateral velocity profiles at the subscale. Heavy particles that move in areas of low horizontal velocity (e.g., near solid surfaces and wake regions) settle closer to the release source as compared to particles in high velocity regions. The macroscopic concentration field of such suspensions is influenced by the ratio between the settling velocity and the subscale distribution of the horizontal velocities. The objective of the study is to isolate and quantify this type of dispersion using controlled flow scenarios. We used a Taylor brush geometry (an array of vertical grooves) to numerically solve the flow. Particles were released from a vertical plane, generating a constant flux. The subscale Eulerian concentration was compiled from simulated trajectories and then spatially averaged to generate macroscopic concentration fields. The results show that this settling-induced dispersion is significant in regions near the source and that it cannot be modeled using Fick's law type of formulations. A parametric investigation shows that the location of the highest dispersion flux is linearly proportional to both the groove spacing and depth. A proposed model that estimates this location is used to evaluate where settling-induced dispersion should not be ignored.

## I. INTRODUCTION

The transport of suspended particles in complex environments is of great importance in many natural and industrial systems. The wide range of natural examples include transport of sediments, bacteria and seeds in fractured rocks (Zvikelsky et al., 2008), in stream gravel beds (Chrysikopoulos and Syngouna, 2014) and in forest canopy (Janhäll, 2015), to name a few. Industrial applications range from water filtration (Woon-Fong, 1998), to enhanced energy recovery (Shiozawa and McClure, 2016) to extraction and separation of chemical, foods or proteins (Lebowitz et al., 2002). The plethora of applications together with the challenging underlying science have motivated intensive particle suspension research, in particular of particle settling and sedimentation (Piazza, 2014).

When using Eulerian descriptions, particle transport is modelled by the advection-dispersion equation (Koch et al., 1989). In complex environments, where fluid velocities vary spatially due to solid obstacles inside the domain, theoretical derivation is obtained by the volume-averaging method. Volume-averaging is obtained by filtering out details at scales smaller than the minimal volume of interest and developing appropriate closure models for dispersion (Whitaker, 1999). The classical work of G.I. Taylor (Taylor, 1953) provides a general closure model for the problem of a solute transport in a non-uniform velocity field, termed the Taylor dispersion. Extension of the Taylor dispersion theory includes examples such as solute dispersion in tubes with arbitrary geometry (Aris, 1956), solute dispersion in a longitudinally varying cross-sections (Gill and Sankarasubramanian, 1970) and particle transport in cylindrical tubes (e.g. Dimarzio and Guttman, 1971; Brenner and Gaydos, 1977). Considerable effort has been devoted to the development of a theoretical framework for the problem of mechanical dispersion resulting from the tortuosity and the variety of particle trajectories within the sub-scale geometries such as in porous media and through vegetation canopies (e.g. Nepf, 1999, Warrick, 2003). However, most of these studies address dissolved solutes, with much fewer works on particle transport; of the latter, the majority of works ignored gravitational effects, assume that the particles are small and neutrally buoyant (Zvikelsky et al., 2008; Piazza et al., 2012) and considered simplified geometries such as a single narrow gap between two parallel horizontal plates (e.g. James and Chrysikopoulos, 2003; Zheng et al., 2009; Meng and Yang, 2016).

In many cases, however, the particles are heavier than the fluid and dispersion models that were developed for solutes or neutrally buoyant particles cannot be applied. This is the case in hydrological problems where the suspended particles are derived from erosion of the solid mineral matrix (e.g. clay particles that are suspended during a flood event). In such cases, gravitational forces cannot be neglected (Zvikelsky et al., 2008) for particles larger than about 1 μm, though recent works highlighted the importance of gravity even for colloids (Piazza et al., 2012). When combined with flow through complex environments, such as cracks, fractures or pores, the effect of both gravity and spatial distribution of the fluid velocity in between the obstacles should be considered.

Heavy particles that move through a complex environment experience a wide variety of processes, including lift force, drag force and advection, settling due to gravity, attachment and detachment, particle-particle interaction and particle-solid interactions, Brownian diffusion, time-dependent inertial forces and more. Out of these many processes, we chose to focus our analysis on the coupling between gravitational settling and horizontal advection of particles within the subscale. Heavy particles that move across areas of low velocity, such as near solid boundaries, settle closer to the release source than those in higher velocity regions, far from the obstacle walls. We name this phenomenon "settling-induced dispersion". The concentration field of suspensions that flow through such complex environments will be determined by the ratio between the settling velocity and the distribution of the stream-wise velocities within the subscale. As we are interested in the fundamental question of how settling impacts particle transport associated with inhomogeneous flow field, we consider the following settings: steady flow in a Taylor brush (Taylor, 1971) made of multiple parallel vertical walls separated by grooves, where the fluid flows inside the grooves and above them (see Fig. 1). This setting is unique as it allows us to obtain an accurate solution for the fully developed flow field between and above the walls, and to follow the trajectories of a large number of particles that are released at the domain inlet. These conditions are designed to investigate the interplay between the two driving mechanisms that lead to settling-induced dispersion: the spatial heterogeneity of the horizontal velocity field and settling due to gravity.

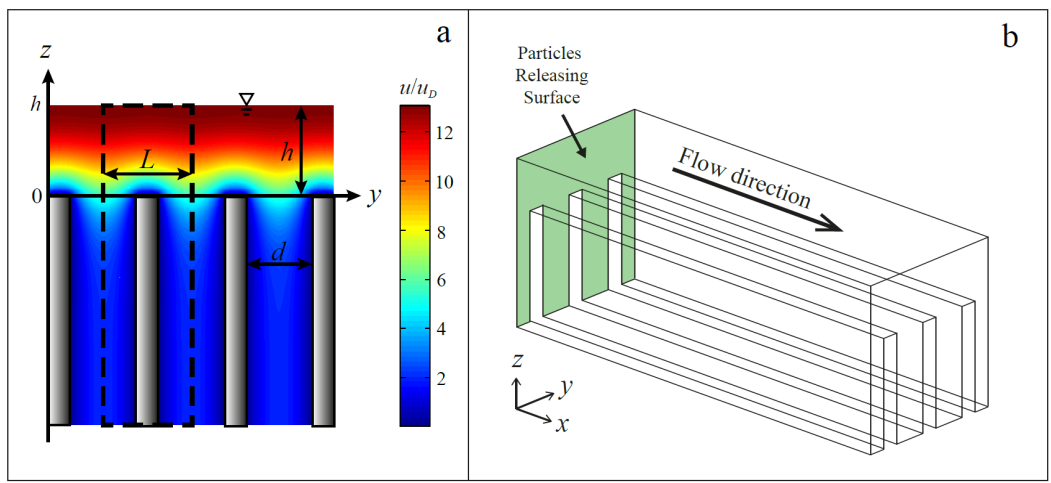

**Fig. 1** The model geometry and the flow field solution. (a) Front view of the Taylor brush configuration superimposed by a solution of the microscopic velocity field (solved for $\phi = 0.75$, $k = 0.05$ cm$^2$). The repeating unit is marked with a dashed line. (b) A schematic 3D view of the brush geometry. Flow is unidirectional in the x direction.

Our model is of relevance to a flow above and within a fracture that is bounded by vertical walls, where particle dispersion is affected by the combination of gravitational settling and horizontal advection. Such flow scenarios occur frequently in both natural and engineered environments, and are thus applicable to a wide range of applications, from water flowing above a fractured soil or rock outcrop (Zvikelsky et al., 2008), to injection of proppants for hydraulic fracturing (Shiozawa and McClure, 2016) and coating of surfaces.

## II. METHODS AND THEORY

The section is divided into four parts; the geometrical configuration of the Taylor brush and the numerical solution of the flow field are described in section A; a definition of the particle Lagrangian motion inside the Taylor brush is given in section B; once the particles trajectories are solved, their Lagrangian description is transformed to Eulerian concentration fields, which are then averaged in space to obtain the macroscopic advection and settling-induced dispersion flux. Section C is, therefore, devoted to the procedure of spatial averaging and the derivation of the macroscopic Eulerian equation with a special emphasis on the derivation and definition of the dispersive flux. Finally, the simulations details and the computation procedures that generate the concentration field are described in section D.

**A. The microscopic geometry and flow solution**

The Taylor brush configuration (shown in Fig. 1) is a synthetic ideal case of spatial heterogeneities in both the subscale, microscopic, domain (the porous-like grooves area) and the macroscopic domain (the free flow region, $z > 0$). Such a simple flow scenario, generated by this geometry, can be used for the separation of settling-induced dispersion and its quantitative analysis presented here. Fluid flows in the $x$ direction within and above the brush grooves while particles move by advection in the $x$ direction and by gravity in the $z$ direction. We characterize the brush geometry by two parameters: the groove density and the hydraulic conductivity. Adopting porous media terminology, they are denoted hereafter as porosity and permeability, defined as $\phi = d/L$ and $k = d^3/(12L) = \phi^3 L^2/12$, where d is the width of the groove and L is the width of the repeating unit. As illustrated in Fig. 1a, the microscopic fluid velocity fields were obtained by solving numerically the 2D Stokes equation in the $y - z$ plane using Comsol Multiphysics, with fluid density, viscosity, and pressure gradient specified as $\rho = 1 \text{ g cm}^{-3}$, $\mu = 0.01 \text{ g cm}^{-1}\text{s}^{-1}$ and $d\langle P\rangle_f = -0.01 \, g \, cm^{-2}s^{-2}$, respectively (Duman and Shavit, 2009; Duman and Shavit, 2010). As the flow is symmetric and periodic in the lateral direction, only half the domain, as marked by the dash line, was solved. No-slip boundary condition was used for all solid surfaces and zero-shear at the water surface and the symmetric boundaries. The height of the free water surface was $h = 1 \, cm$. It was found that this water height is appropriate as the particle concentration above the interface becomes uniform with $z$ and that choosing a thicker water layer has no effect on the study conclusions. The microscopic velocity solution obtained here was later averaged with respect to $y$ in order to calculate the macroscopic velocity profile.

**B. Particle motion**

Suspended particles are subjected to a wide variety of forces and transport mechanisms. As the motivation of the study is to investigate the effect of settling on the dispersion of particles, we chose to focus the simulations on the relevant forces. First, only dilute mixtures were considered and therefore particles are assumed to exhibit a one-way coupling with the fluid flow, i.e. particle motion depends on the fluid flow field but not vice versa. The effect of lift force, unsteady forces (added mass and Basset history effect) and Brownian diffusion were assumed to be negligible relative to that of drag and gravity. Under these assumptions, the Lagrangian momentum equation of the particle motion is reduced to the following form:

$$\frac{dU_{pi}}{dt} = \frac{u_i(\boldsymbol{x_p}, t) - U_{pi}}{\tau_p} + g\delta_{i3} \tag{1a}$$

$$\frac{dx_{pi}}{dt} = U_{pi} \tag{1b}$$

Where $U_{pi}$ and $u_i$ ($i = 1,2,3$) are the particle Lagrangian velocity along its trajectory and the fluid's Eulerian velocity respectively. The particle location $\boldsymbol{x_p}$ is a vector and is therefore marked in bold. The particle coordinates $x_{p1} = x_p$, $x_{p2} = y$ and $x_{p3} = z_p$; particle velocity components $U_{p1} = U_p$, $U_{p2} = V_p$ and $U_{p3} = W_p$; and the fluid's velocity components $u_1 = u$, $u_2 = v$ and $u_3 = w$, are aligned along $x$, $y$, and $z$ respectively. $\delta_{ij}$ is a Kronecker delta, $t$ is time and $g$ is the gravitational acceleration, which acts only in the vertical direction ($i = 3$). The particle relaxation time in laminar flow is given by $\tau_p = \rho_p d_p^2/(18\mu)$, where $\rho_p$ is the particle density and $d_p$ is its diameter (Hinds, 1999).

A solution of Eq. 1 shows an exponential response of the particle velocity that decays towards the fluid's velocity. Here we assume that the decay occurs fast enough to obtain "full-drag", which means that the particle acquires the fluid velocity instantly. Hence, for the current case, the numerical scheme of Eq. 1 is written as:

$$x_{pi}(t + \Delta t) = x_{pi}(t) + [u_i(\boldsymbol{x_p}, t) + w_g\delta_{i3}]\Delta t \tag{2}$$

where $w_g$ is the gravitational Stokes settling velocity, $w_g = \tau_g g$.

The simulations of particle transport in the brush configuration include the release of a large number of particles from the $y - z$ plain at $x = 0$ that serves as the particle source plain (see shaded surface in Fig. 1b). The particles are released at a constant flux, with an initial velocity that is equal to the local microscopic fluid velocity $u$ in $x$ and to the Stokes settling velocity $w_g$ in $z$. The initial velocity in the lateral $y$ direction is zero ($V_p = v = 0$) and stays zero for the entire simulation. The trajectory of each particle is calculated using Eq. 2 until it reaches either the top of a wall or the bottom of the domain (at $z = -2.5h$). Both boundaries are treated as a perfect sink, such that a particle that settles on them is absorbed by the wall and cannot be re-suspended.

## C. The macroscopic Eulerian transport equation of suspended particles

The derivation of the macroscopic equation begins with the microscopic, subscale, mass balance equation,

$$\nabla \cdot (C\boldsymbol{u_p}) = 0 \tag{3}$$

where $C$ is the microscopic concentration of the particles phase and $\boldsymbol{u_p}$ is the microscopic Eulerian velocity vector of that phase. In general, a solution of Eq. 3 provides a steady state description of the microscopic concentration field while neglecting molecular diffusion. The macroscopic equation is derived by averaging Eq. 3 in space as follows:

$$\langle \nabla \cdot C\boldsymbol{u_p} \rangle = \nabla \cdot \langle C\boldsymbol{u_p} \rangle + \frac{1}{V}\int_A (C\boldsymbol{u_p}) \cdot \hat{n} dA = 0 \tag{4}$$

The angle brackets $\langle \ \rangle$ represent the spatial superficial averaging over $V$, the representative elementary volume (REV), $A$ is the total area of the fluid/solid interface and $\hat{n}$ is the unit normal vector of $A$ pointing into the solid. The integral term in Eq. 4 is a sink term representing the collection of particles on the solid surfaces (Whitaker, 1999).

Next, the term $\langle C\boldsymbol{u_p} \rangle$ is treated through spatial decomposition:

$$C = \langle C \rangle_f + \tilde{C}; \qquad \boldsymbol{u_p} = \langle \boldsymbol{u_p} \rangle_f + \tilde{\boldsymbol{u}}_p \tag{5}$$

where $\langle C \rangle_f$ is the intrinsic average (over the fluid fraction of the averaging volume), and $\tilde{C}$ and $\tilde{\boldsymbol{u}}_p$ are the intrinsic spatial fluctuations of the concentration and velocity. The relationship between the superficial and intrinsic averages is given by $\langle \ \rangle = \phi \langle \ \rangle_f$ where $\phi$ is the local porosity. The result of the spatial decomposition is as follows:

$$C\boldsymbol{u_p} = \phi^{-1}\langle C \rangle \langle \boldsymbol{u_p} \rangle + \langle \tilde{C}\tilde{\boldsymbol{u}}_p \rangle \tag{6}$$

By combining Eqs. 4 and 6 we write the macroscopic mass balance equation:

$$\nabla \cdot \left( \phi^{-1}\langle C \rangle \langle \boldsymbol{u_p} \rangle \right) + \nabla \cdot \langle \tilde{C}\tilde{\boldsymbol{u}}_p \rangle + S = 0 \tag{7}$$

where $S$ replaces the integral term in Eq. 4. We can now use Eq. 7 to solve the transport problem for the Taylor brush configuration. Under the assumption of "full-drag", the streamwise velocity of the particles and the fluid velocity are identical ($\langle u_p \rangle = \langle u \rangle$ and $\tilde{u}_p = \tilde{u}$); the vertical particles velocity is constant ($\langle w_p \rangle = w_g$ above the grooves domain, $\langle w_p \rangle = \phi w_g$ inside the grooves domain and $\tilde{w}_p = 0$ everywhere). The lateral velocity is zero ($\langle v_p \rangle = 0$ and $\tilde{v}_p = 0$). The source term $S$ is zero everywhere except for $z = 0$, where it represents particles accumulation on the horizontal upper surface of the walls ($S = C(z = 0)w_g(1 - \phi)\Delta z^{-1}$, where $\Delta z$ is the REV vertical dimension, see section 2.4).

Eq. 7 is therefore reduced inside and above the grooves domain to:

$$\langle u \rangle \frac{\partial \langle C \rangle}{\partial x} + w_g \frac{\partial \langle C \rangle}{\partial z} + \frac{\partial}{\partial x}\langle \tilde{C}\tilde{u} \rangle = 0 \qquad \text{for} \quad z > 0 \tag{8a}$$

$$\langle u \rangle \frac{\partial \langle C \rangle}{\partial x} + \phi w_g \frac{\partial \langle C \rangle}{\partial z} + \phi \frac{\partial}{\partial x} \langle \widetilde{C} \widetilde{u} \rangle = 0 \qquad \text{for} \quad z < 0 \tag{8b}$$

Note, that for the transition from Eq. 7 to Eq. 8, $\langle C \rangle \partial \langle u \rangle / \partial x$ is zero since the velocity does not change in the $x$ direction. The last term of Eq. 8 is the $x$-component dispersive flux which is the subject matter of this study. In order to solve Eq. 8, a closure model is needed for the dispersive flux in terms of the mean concentration $\langle C \rangle$. It will be shown later that the use of a diffusion model is, unfortunately, not suitable.

### D. The simulation parameters and the averaging procedure

Numerical simulations were used to generate particle trajectories for seven values of permeability (0.001, 0.005, 0.01, 0.02, 0.05, 0.075 and 0.1 cm$^2$) and two values of porosity (0.75 and 0.9). Together with the hydraulic conditions defined earlier (section 2.1) this range of values generates a Darcy flux between 0.001 and 0.1 cm s$^{-1}$, as often found in large fractures (e.g. Grisak and Pickens, 1981; Wan et al., 1996). The simulations were performed with a single particle terminal velocity of $w_g = -0.01\ cm\ s^{-1}$. Note that under these flow conditions the Stokes length scale is $w_g \tau_p = 10^{-8} h$, indicating that the "full-drag" assumption is valid. Such a terminal velocity represents large clay particles or small silt particles of diameter around 10 μm (e.g., Wentworth, 1922). In the simulations, particles were released from the vertical $y - z$ plane at $x = 0$ (see Fig. 1b) with a continuous flux ($F_0$) of 2.4·10$^6$ particles per second from the surface of one repeating unit (marked with a dashed line in Fig. 1a). The simulations lasted until a steady state was obtained, i.e., the number of particles in the domain did not change with time. The number of particles becomes steady because the time it takes to the "most persistent" particle to be deposited and hence leave the domain has been exceeded. From this moment on the number of particles entering the domain through the upstream section is equal to the number of particles that leaves the domain, hence a constant number of particles is reached within the domain.

The computations of the Lagrangian trajectories result in a data set of locations and velocities along the trajectory of each individual particle. The microscopic concentration field was then calculated by binning the particles into a Cartesian grid (based on their locations), summing the number of particles in each grid cell and dividing the sum by the cell volume (Δx×Δy×Δz). This process produces the three-dimensional microscopic (sub-scale) concentration $C$ (number of particles per unit volume). Since the concentration field is changing in all three dimensions, choosing the volume of the binning cell is a compromise between a cell that is large enough to contain a meaningful number of particles and small enough when compared to the distance between the walls ($d$ in Fig. 1) and the height of the free water region, $h$. A detailed analysis has been obtained to find a cell-volume that is sufficiently large to avoid noisy concentration profiles and at the same time sufficiently small to avoid undesired filtering of concentration variations. A choice of Δy = d/75, $\Delta z = h/115$, and $\Delta x = 0.2h - 0.3h$ (depending on the porosity and permeability for each case) was found to satisfy these constraints.

Once the microscopic concentration $C$ was calculated, spatial averaging was applied to generate the macroscopic concentration field, $\langle C \rangle$. This averaging procedure involves a dimensions-choice for the representative elementary volume (REV) and results in a mean and fluctuations values. The size of the REV is compared here to the macroscopic dimensions of the problem. The REV in the $y$ direction was chosen as the entire width of the repeating unit ($L$). Following the canopy flow literature (e.g., Raupach and Shaw, 1982), the size of the REV in the $x$ and $z$ directions was chosen to be as small as possible to avoid smoothing of the macroscopic concentration change, and therefore it is equal to the binning cell dimensions; $\Delta x$ and $\Delta z$.

## III. RESULTS

### A. Mean concentration profiles

An example of mean concentration profiles is shown in Fig. 2. The profiles are presented as a function of $x$ for six different $z$ heights and for a porosity of 0.75 and a permeability of 0.02 cm². The concentration values are normalized by a reference value $C^* = F_0/u_D$, where $F_0$ is the source release flux, and $u_D$ is the Darcy velocity (the mean fluid velocity deep inside the grooves-porous-like domain, $\langle u \rangle|_{z \to -\infty}$). Note, that the reference concentration is not affected by the choice of the REV size.

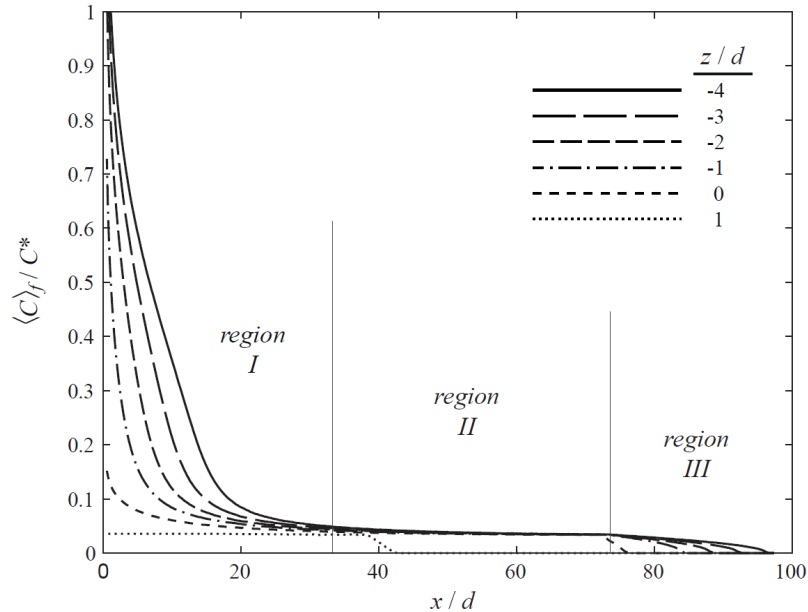

**Fig. 2** Mean concentration profiles at several z locations ($\phi = 0.75$, $k = 0.02$ cm²). Three regions are marked for the $z = -4d$ curve: region $I$ with a rapid decrease of concentration; region $II$ with a stable concentration; and region $III$ where the concentration decays to zero. Regions are marked by two separating vertical lines.

The spatial distribution of particles can be classified by three distinct regions. A rapid decrease in particle concentration that occurs near the source (region $I$); a quasi-constant concentration in the middle section (region $II$); and particles that sink and exit the domain at the far end (region $III$), leaving no particle further downstream. The reason for the high loss of particles near the source is due to particles that are released inside the grooved domain and settle earlier as a result of the low velocities near the walls. The length of region $I$ (in the $x$ direction) is increasing as we go deeper into the grooved domain because of particles that arrive from upper layers. The reason for the stable concentration along region II is the constant flux of particles that were initially released above the interface ($x = 0$, $z > 0$) and serves, along this region, as a source of particles. This source is depleted in region III and as a result the concentration decreases to zero towards the end of this region ($x/d \approx 100$). This concentration behavior is a result of the combined effect of heterogeneities in both the sub-scale and the macroscopic domains. Particles in fractured media or Karst formations often move from large cavities into narrow cavities and vice-versa. A similar scenario often occurs within gravel-beds of rivers and streams. This heterogeneity is represented, in part, by the coexistence of an open flow region and the grooves region. Particles transport in a hypothetical infinite domain (in the $z$ direction) that has no free flow region, will result in a single concentration value. As in the case of region $II$, the reason for this single concentration is a balance between a constant flux of particles that arrive from above and the flux of particles that continue to sink to lower layers. Finally, the only result that is affected by the choice of $h$ (= 1 cm) is the length of region $II$ as high $h$ will increase the length of region $II$.

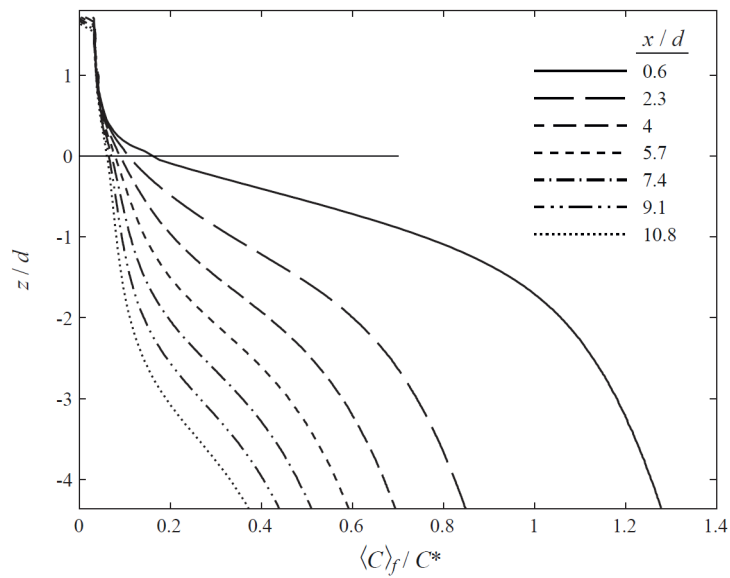

**Fig. 3** Vertical mean concentration profiles at several x locations ($\phi = 0.75$, k = 0.02 cm$^2$). The top of the walls is indicated by a horizontal solid line ($z = 0$).

Fig. 3 shows mean concentration profiles as a function of $z$ at several $x$ locations in region $I$. The curves were plotted at $x/d$ locations that were uniformly spread ($\Delta x/d = 1.7$), showing the rapid decay of concentration along $x$ while keeping, in this region, an almost constant concentration above the interface ($z > 0.5d$). Looking at the most upstream profile ($x = 0.6d$) we see that immediately below the top of the walls ($z = 0$), the concentration rapidly increases and then, when moving deeper into the grooves domain, the profile tends towards a constant value. The section with the rapid concentration increase is influenced by settling and hence is located at lower and lower heights as we move downstream, away from the source. The uniform concentration above the interface indicates that the choice of water height was appropriate and increasing the water thickness above the interface has no effect within this region.

**B. The dispersive flux**

The choice of the brush configuration provides an ideal case study by which the dispersion flux $\langle \tilde{C} \tilde{u} \rangle$ can be directly calculated in order to investigate the effect of coupling between the settling velocity and the microscopic velocity field. The results of the dispersion flux as a function of $x$ at five heights are shown in Fig. 4. The dispersion flux near the source is highly negative; it increases steeply, turns positive and reaches a maximum (at $x_m$) before decaying towards zero in the downstream regions. Far enough from the source, settling-induced dispersion is negligible and the streamwise advection is balanced by vertical advection ($\langle u \rangle \partial \langle C \rangle / \partial x + \phi w_g \partial \langle C \rangle / \partial z$). However, the area where dispersion cannot be neglected is significant, especially deep inside the porous-like region. The magnitude of the dispersion flux decreases with increasing $z$ and then disappears (by definition) above the interface (see the curve for $z/d = 1$).

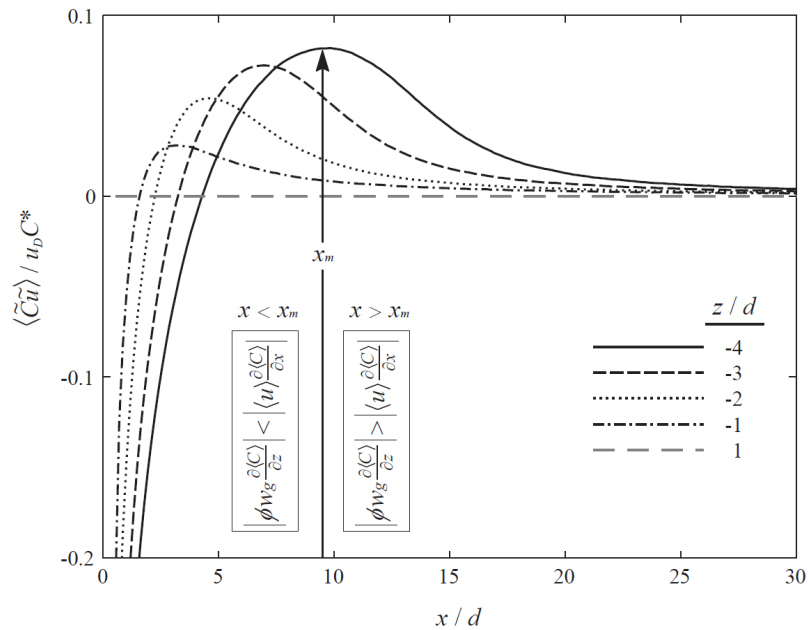

**Fig. 4** The dispersion flux as a function of $x$, calculated at several $z$ locations ($\phi = 0.75$, k = 0.02 cm$^2$). The value of $x_m$ (location of the maximum dispersion flux) and the subsequent relation between the horizontal and vertical advective fluxes are illustrated for the $z/d = -4$ case.

In order to explain the results of Fig. 4 we generated in Fig. 5 a plot of the concentration fluctuations, $\tilde{C}$, across a single groove. The three curves shown in Fig. 5 represent different $x$-locations while all three were taken at the same depth ($z/d = -4$). Every point in Fig. 4 was generated by applying a correlation between concentration profiles such as those shown in Fig. 5 and the velocity fluctuation, $\tilde{u}$ (also shown in Fig. 5 as a thick solid blue curve). The negative dispersion that always appears near the source (i.e., at low $x/d$ values in Fig. 4) is explained in Fig. 5 by the correlation of the $\tilde{u}$ curve and the $\tilde{C}$ curve at $x = d$. Since these two curves have opposite signs across the groove, the correlation is negative. Since the magnitude of $\tilde{C}$ is larger than in the downstream curves, the resulting dispersion flux near the source is highly negative. Next, Fig. 4 shows that the dispersion flux is becoming less negative with increasing x. This behavior is explained in Fig. 5 by showing that since the particles near the walls settle and disappear, the magnitude of $\tilde{C}$ is decreasing and the location of its maximum value moves towards the center of the groove. This continues until $\langle \tilde{C} \tilde{u} \rangle$ reaches a maximum value at $x_m$. Fig. 5 shows that at $x = x_m$, $\tilde{C}$ and $\tilde{u}$ have the same sign which explains the positive values of the dispersion fluxes. Further downstream (Fig. 5, $x = 15d$), the sign of the two curves is mostly the same but the magnitude of $\tilde{C}$ decreases (together with the mean concentration) and the dispersion flux becomes zero.

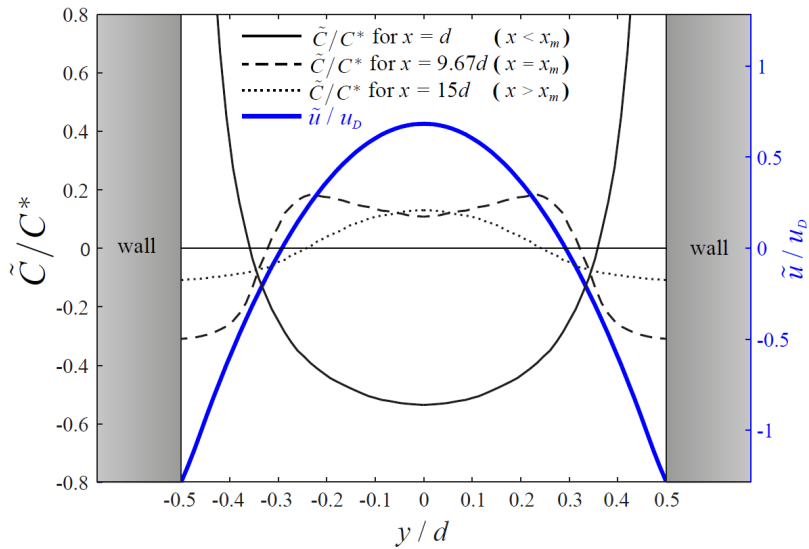

**Fig. 5** Concentration and velocity fluctuation profiles in the groove between the brush walls at $z = -4d$ ($\phi = 0.75$, $\mathrm{k} = 0.02\ \mathrm{cm}^2$).

The common use of Fick's law as a model for dispersion flux (Bear, 1972) means that it is scaled with the gradient of the mean concentration,

$$\langle \tilde{C} \tilde{\boldsymbol{u}} \rangle = -D_{disp} \nabla \langle C \rangle, \tag{9}$$

where $\mathrm{D_{disp}}$ is the dispersion coefficient. While the traditional approach is to model dispersion as a Fickian process, the current results show that $\langle \tilde{C} \tilde{u} \rangle$ has both positive and negative values (as shown in Fig. 4), a result that cannot be represented by a Fickian formulation without a change in the sign of the concentration gradient. Figs. 2 and 3 show that the concentration decreases with $x$ and increases with depth such that both $\partial \langle C \rangle / \partial x$ and $\partial \langle C \rangle / \partial z$ are always negative. The conclusion is that settling-induced dispersion, at least for the current problem, cannot be modeled using Fick's law.

## C. The location of the maximum dispersion flux ($x_m$)

The location $x_m$, where the dispersion flux reaches its maximum value (Fig. 4), has a special meaning. Dispersion on the left side of the maximum ($x < x_m$) serves, by analogy, as a sink term and vice versa; dispersion on the right side of the maximum ($x > x_m$) serves as a source term. Such sink/source maps can be drawn along the $x - z$ vertical plane. The first term of Eq. 8 (the contribution of the advective flux in $x$) is always negative since $\langle u \rangle$ is positive but $\partial \langle C \rangle / \partial x$ is always negative (see Fig. 2). The second term (the contribution of the advective flux in $z$) is always positive since both $w_g$ and $\partial \langle C \rangle / \partial z$ are negative. This sign analysis means that at a given point, the first term adds particles (as in source terms) and the second removes particles (sink). Obviously, Fig. 4 shows that until the maximum point of $\langle \widetilde{C} \widetilde{u} \rangle$ is reached, its derivative is positive $\partial \langle \widetilde{C} \widetilde{u} \rangle / \partial x > 0$, which means that at any given point in this region the dispersion term in Eq. 8 behaves as a source, just like the second term of Eq. 8. It means that for $x < x_m$ the advective flux balance is larger in $x$ than in $z$. Downstream from $x_m$, $\partial \langle \widetilde{C} \widetilde{u} \rangle / \partial x$ is negative and therefore the balance reverses. This role of dispersion is illustrated schematically in Fig. 4 and summarized in Table 1.

**Table 1.** The sign of the three terms in Eq. 8.

|  | $x < x_m$ | $x > x_m$ |
|---|---|---|
| $\langle u \rangle \partial \langle C \rangle / \partial x$ | − | − |
| $\phi w_g \partial \langle C \rangle / \partial z$ | + | + |
| $\phi \partial \langle \widetilde{C} \widetilde{u} \rangle / \partial x$ | + | − |

In addition to the change in sign, $x_m$ separates between the region where the contribution of dispersion is large versus the region where its contribution becomes small and negligible. It is, therefore, important to be able to predict the location of $x_m$. In Fig. 4 we noticed that $x_m$ is increasing with $z$. Fig. 6 presents the effect of permeability on the dispersion flux at $z = -2.25 \, cm$ and shows that $x_m$ is increasing with the permeability and hence the distance in which the dispersion is significant becomes longer ($0 < x < x_m$).

We therefore plot Fig. 7 in which we show the behavior of $x_m$ as a function of permeability. It is shown that $x_m$ is a linear function of $k$ (Fig. 7a) and $z$ (Fig. 7b). As was demonstrated in Fig. 5, the dispersion flux is significant when particles are still present in the domain, especially close to the walls. We therefore postulate that a maximum in the dispersion flux will occur at a location that is correlated with the time it takes for particles to settle and leave the domain, $\hat{x}_m = \langle u \rangle t_0$, where $t_0$ is the time it takes for a particle to reach a specific $z$, i.e., $t_0 = -z/w_g$.

As an approximation we use the mean velocity inside the brush, calculated by the Poiseuille solution for the Taylor brush geometry $\langle u \rangle = -d^2 (d\langle P \rangle_f/dx)/(12\mu)$, such that a model that estimates the maximum location is derived as:

$$\hat{x}_m = \langle u \rangle t_0 = -\langle u \rangle \frac{z}{w_g} = \frac{d^2}{12\mu} \frac{d\langle P \rangle_f}{dx} \frac{z}{w_g} \tag{10}$$

and since $d^2 = 12k/\phi$,

$$\hat{x}_m = \frac{d\langle P \rangle_f}{dx} \frac{kz}{\mu \phi w_g} \tag{11}$$

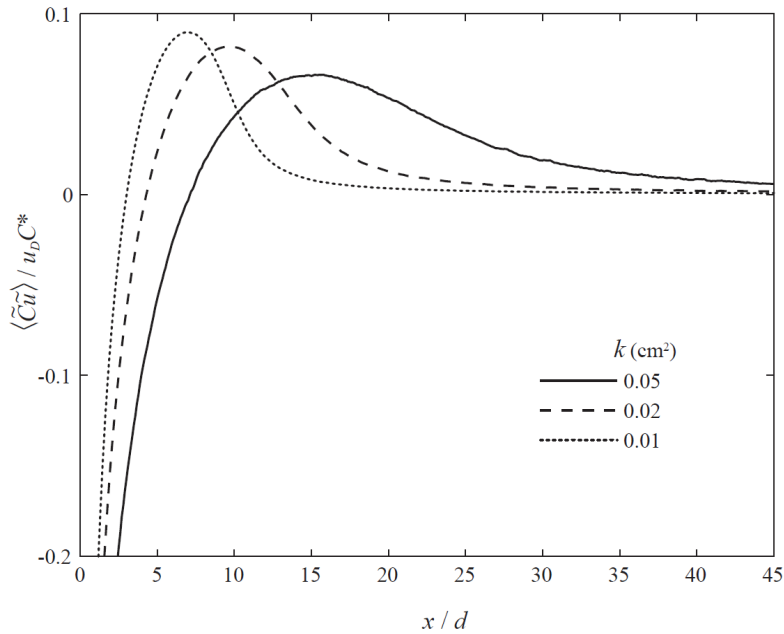

**Fig. 6** Dispersion flux profiles for three permeability values ($\phi = 0.75$, $z = -2.25\ cm$). Note that $u_D$ and $d$ are different for each permeability: for $k = (0.01, 0.02, 0.05)$, $d = (0.4, 0.57, 0.9)$ and $u_D = (0.01, 0.02, 0.05)$.

Eq. 11 indicates that the relationship between the estimated maximum location and the permeability is indeed linear. It is also linear with $z$ and with the inverse of $\phi$. The linear behavior as a function of $k$ is validated using the simulation results shown in Fig. 7a. As shown, Eq. 11 overestimates the location of the maximum value, especially in the deeper layers. Note, that the overestimation appears where the mass balance (Eq. 8) is almost unaffected by the dispersion term.

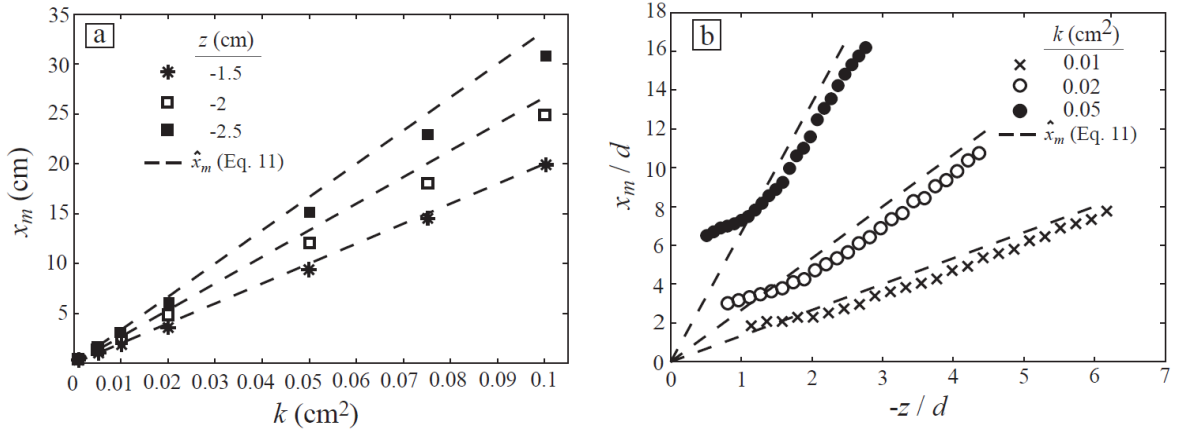

**Fig. 7** The location of the maximum value of the dispersion flux (a) as a function of permeability and (b) as a function of $z$ ($\phi = 0.75$). The symbols represent the results of the simulation and the dashed lines represent the estimation of the peak location using Eq. 11.

Finally, we validate the relation between the maximum location and $z$. Fig. 7b shows that Eq. 11 serves as a good estimate except for the regions near the interface, where $\langle u \rangle$ is large, however, the contribution of the dispersive flux at these heights is less significant (see Fig. 4). Eq. 11 predicts again a slight over estimation but still provides a good approximation of the simulations results.

## IV. SUMMARY AND CONCLUSION

Settling-induced dispersion depends on the microscopic distribution of the horizontal velocities since heavy particles that are found in regions of low velocity tend to sink earlier than those in regions of high velocity. In the current study we isolated and then analyzed this dispersion mechanism by using a Taylor brush which served as a simplified model medium of systems that contain vertical fractures and voids. Simulations of heavy particles were obtained by calculating their trajectories, which allowed a detailed calculation of the microscopic concentration field. This calculated concentration field was averaged in space and then used to estimate the dispersion flux. It was found that settling-induced dispersion is important in regions near the source and cannot be neglected there. Further downstream, it decays and only advection takes part. It was shown that the dispersion flux is not proportional to the concentration gradient and therefore cannot be modeled using Fick's law. Based on the study results, the section along which settling-induced dispersion is important ($0 < x < x_m$) was quantified. It was found that the value of $x_m$ is linearly proportional to both the permeability and depth. We then proposed a model that predicts $x_m$ which proved to generate a good agreement with the simulations. Hence, when the flow geometry (groove spacing and conductivity) and the interface location are known, and the flow parameters are given, an estimation tool is now available for the region in which settling-induced dispersion is to be considered. It still remains a challenge to find an alternative closure model for this dispersion flux and to study the combined effect when other types of dispersion exist.

The study objective was to demonstrate the possible influence of settling-induced dispersion and to analyze a flow scenario which is unidirectional and characterized by a sub-scale geometry that is as simple as possible. We claim that this type of dispersion exists in any complex geometrical configuration and when the flow is neither unidirectional nor steady. Since complex scenarios are more difficult to solve and since they are characterized by a large number of geometrical, flow and particle parameters and hence a large number of combinations, we focused the study on the case analyzed here. The study presents a theoretical case by which the mechanism of settling-induced dispersion is now better understood. Future studies should expand beyond the conditions tested here and examine turbulent velocity profiles, geometries that generate lateral and vertical sub-scale velocities, the competition between attachment on the solid surfaces and dispersion and to include other macroscopic flows such as infiltration and upwelling.

**ACKNOWLEDGMENTS**

This research was supported by the Israeli Minister of Science, Technology and Space (grant #3-12478). RH also acknowledges partial support from the United States-Israel Binational Science Foundation (#BSF-2012140).

**REFERENCES**

1. Aris, R.: On the dispersion of a solute in a fluid flowing through a tube, *Proc. R. Soc. A* **235**(1200), 67-77 (1956).

2. Bear, J., *Dynamics of fluids in porous media* (Elsevier, New York, 1972).

3. Brenner, H. Gaydos, L. J.: The constrained Brownian movement of spherical particles in cylindrical pores of comparable radius models of the diffusive and convective transport of solute molecules in membranes and porous media, *J. Colloid Interface Sci.* **58** (2) 312–356 (1977).

4. Chrysikopoulos, C. V. and Syngouna, V. I.: Effect of gravity on colloid transport through water-saturated columns packed with glass beads: Modeling and experiments, *Environ. Sci. Technol.* **48**, 6805–6813 (2014).

5. Dimarzio, E.A. Guttman, C.M.: Separation by flow and its application to gel permeation chromatography, *J. Chromatogr. A* **55**(1), 83–97 (1971).

6. Duman, T. and Shavit, U.: An apparent interface location as a tool to solve the porous interface flow problem, *Transp. Porous Med.* **78**, 509–524 (2009).

7. Duman, T. and Shavit, U.: A solution of the laminar flow for a gradual transition between porous and fluid domains, *Water Resour. Res.* **46**, W09517 (2010).

8. Gill, W. N., Sankarasubramanian, R., Exact analysis of unsteady convective diffusion, *Proc. R. Soc. Lond. A* **316**, 341–350 (1970).

9. Grisak, G. E. and Pickens, J. F.: An analytical solution for solute transport through fractured media with matrix diffusion, *J. Hydrol.* **52**(1), 47-57 (1981).

10. Hinds, W. C.: *Aerosol technology: Properties, behavior, and measurement of airborne particles*, 2nd ed. (Wiley, New York, 1999).

11. Janhäll, S.: Review on urban vegetation and particle air pollution – Deposition and dispersion, Atmos. Environ., 105 130-137 (2015).

12. James, S. C. and Chrysikopoulos, C. V.: Effective velocity and effective dispersion coefficient for finite-sized particles flowing in a uniform fracture. *J. Colloid Interface Sci.* **263**, 288–295 (2003)

13. Koch, D.L., Cox, R.G., Brenner, H. and Brady, J.F.: The effect of order on dispersion in porous media, *J. Fluid Mech.* **200**(1), 173-188 (1989).

14. Lebowitz, J., Lewis, M. S., Schuck, P.: Modern analytical ultracentrifugation in protein science: A tutorial review, *Protein Sci*. **11**, 2067–79 (2002).

15. Meng, X., Yang, D.: Determination of dynamic dispersion coefficient for particles flowing in a parallel-plate fracture, *Colloids and Surfaces A,* **509**, 259–278 (2016).

16. Nepf, H.: Drag, turbulence, and diffusion in flow through emergent vegetation, *Water Resour. Res.* **35**(2), 479-489 (1999).

17. Piazza, R., Buzzaccaro, S. and Secchi, E.: The unbearable heaviness of colloids: Facts, surprises, and puzzles in sedimentation, *J. Phys: Condens. Matt*. **24**, 284109 (2012).

18. Piazza, R., Settled and unsettled issues in particle settling, *Rep. Prog. Phys*. **77**, 056602 (2014).

19. Raupach, M.R., and Shaw, R.H.: Averaging procedures for flow within vegetation canopies, *Boundary-Layer Meteorol*., **22**(1), 79-90 (1982).

20. Shiozawa, S., and McClure, M.,: Simulation of proppant transport with gravitational settling and fracture closure in a three-dimensional hydraulic fracturing simulator*, J. Pet. Sci. Eng.* **138**, 298-314 (2016).

21. Taylor, G. I.: Dispersion of soluble matter in solvent flowing slowly through a tube, *Proc. R. Soc. A* **219**, 186-203 (1953).

22. Taylor, G. I.: A model for boundary condition of a porous material, *J. Fluid Mech.* **49**, 319–326 (1971).

23. Wan, J., Tokunaga, T.K. Tsang, C.F. and Bodvarsson, G.S.: Improved glass micromodel methods for studies of flow and transport in fractured porous media, *Water Resour. Res.* **32**(7), 1955-1964 (1996).

24. Warrick, A.W., Soil water dynamics, (Oxford University Press, Oxford, 2003).

25. Wentworth, C.K.: A scale of grade and class terms for clastic sediments, *J. Geol.* **30**(5), 377-392 (1922).

26. Whitaker, S., *The method of volume averaging* (Kluwer, Dordrecht, 1999).

27. Woon-Fong L. W., *Industrial Centrifugation Technology* (New York: McGraw-Hill, 1998).

28. Zheng, Q., Dickson, S.E., Guo, Y.: Differential transport and dispersion of colloids relative to solutes in single fractures, *J. Colloid Interface Sci*. **339**(1), 140–151 (2009).

29. Zvikelsky, O., Weisbrod, N., and Dody, A.: A comparison of clay colloid and artificial microsphere transport in natural discrete fractures*. J. Colloid Interface Sci*. **323** ,286–292, (2008).